\documentclass[a4paper,11pt]{article}
\pdfoutput=1
\usepackage{jinstpub}

\usepackage{lineno}
\usepackage{xspace}

\newcommand\inquotes[1]{\lq#1\rq}
\newcommand\Fig     [1]{Figure~\ref{#1}\xspace}

\newcommand\fig     [1]{Fig.~\ref{#1}}

\def\nm{\mathrm{\,   nm}}
\def\um{\mathrm{\,\mu m}}
\def\mm{\mathrm{\,   mm}}
\def\cm{\mathrm{\,   cm}}

\def\ps{\mathrm{\,   ps}}

\def\us{\mathrm{\,\mu s}}

\def\Ohm {\mathrm{\, \Omega}}

\def\V   {\mathrm{\,    V}}

\def\mA  {\mathrm{\,  m A}}
\def\Amp {\mathrm{\,    A}}

\def\MHz{\mathrm{\,MHz}}
\def\GHz{\mathrm{\,GHz}}

\def\W  {\mathrm{\,    W}}

\def\MBq {\mathrm{\,MBq}}

\def\PETAT  {PETAT\xspace}
\def\PETATa {PETAT1\xspace}
\def\PETATb {PETAT2\xspace}

\newcommand {\Figure}[4] {
 \begin{figure}[!t]
 \centering
 \includegraphics[width=#4\linewidth]{figures/#1}
 \vskip-2mm
 \caption{#3} \label{#2}
 \end{figure}
}

\title{\boldmath \PETAT\ - An ASIC for Simple and Efficient Readout of Large PET Scanners}

\author{P. Fischer}
\author{and T. Kerschenbauer}
\author{and M. Ritzert}
\affiliation{Institute for Computer Engineering, Heidelberg University, Germany}
\emailAdd{peter.fischer@ziti.uni-heidelberg.de}

\abstract{
Modern PET scanners based on scintillating crystals use solid state photo detectors for light readout. The small area of these devices is beneficial for spatial resolution, but also leads to a large number of electronic channels to be read out, mostly by application specific integrated circuits (ASICs) containing amplification, noise reduction, hit finding, time stamping and amplitude measurement. Although each ASIC provides up to $\approx 64$ channels, a large number of chips is required with the need for auxiliary electronic components like voltage regulators or FPGAs for control and data readout. The FPGAs in turn often require multiple supply voltages and configuration infrastructure, so that PCBs get complicated, cumbersome and power-hungry, in addition to the significant power requirement of the front-end ASICs.
We address this issue in the latest generation of our PETA readout ASIC for SiPMs by a simplified control scheme and, in particular, by a hierarchical serial data readout which does not require any additional FPGA. In addition, it provides a time-sorted stream of hit data, allowing early on-detector data reduction and hit pre-processing like the removal of hits with no coincident partner.
The simplicity of this readout facilitates a supply scheme where power/ground of multiple ASICs are connected in series instead of the standard parallel connection.
This \inquotes{serial-powering} approach can reduce supply current (while increasing overall supply voltage) so that voltage drop issues in the supply are alleviated.}


\hypersetup{pdfauthor={Peter Fischer et al.},pdftitle={Efficient PET Readout}}

\begin{document}
\maketitle 
	
\section{Motivation}

Modern PET scanners use an increasingly high number of electronic channels to improve position- and time resolution, or to increase the sensitivity by a larger
solid angle coverage (see for instance \cite{largescanner}). This leads to at least two challenges: a) The hit data generated in the large number of different front-end chips must be collected and sent to the data acquisition module. As illustrated in \fig{fig:classicalsystem}, this task is typically done in FPGAs (field programmable gate arrays) in the vicinity of the front-end chips, typically one FPGA per physical \inquotes{module} unit. The FPGAs, in turn, require substantial auxiliary circuitry for power supply (often multiple voltages), for providing a configuration bit stream, and for data readout. This adds significant complexity, space requirement, cost and power consumption. b) The large number of chips leads to a significant power need. As the supply voltages of the chips tend to decrease with smaller transistor feature sizes ($1-2\V$), the supply currents often increase. In order to limit resistive voltage drops at high overall currents, cable resistivity must be kept very low and clumsy cables with large cross sections are unavoidable. Lower currents in the supply cables can be obtained by switched-mode step-down converters close to the front-end chips \cite{safirdcdc}, adding complexity and the risk for electromagnetic interference (in particular in PET-MRI systems) or poor supply voltage quality. Using local linear regulators to compensate for voltage drops again adds components and leads to an additional power dissipation. Another significant challenge in PET system design is the removal of the dissipated power by an efficient cooling system.

\Figure{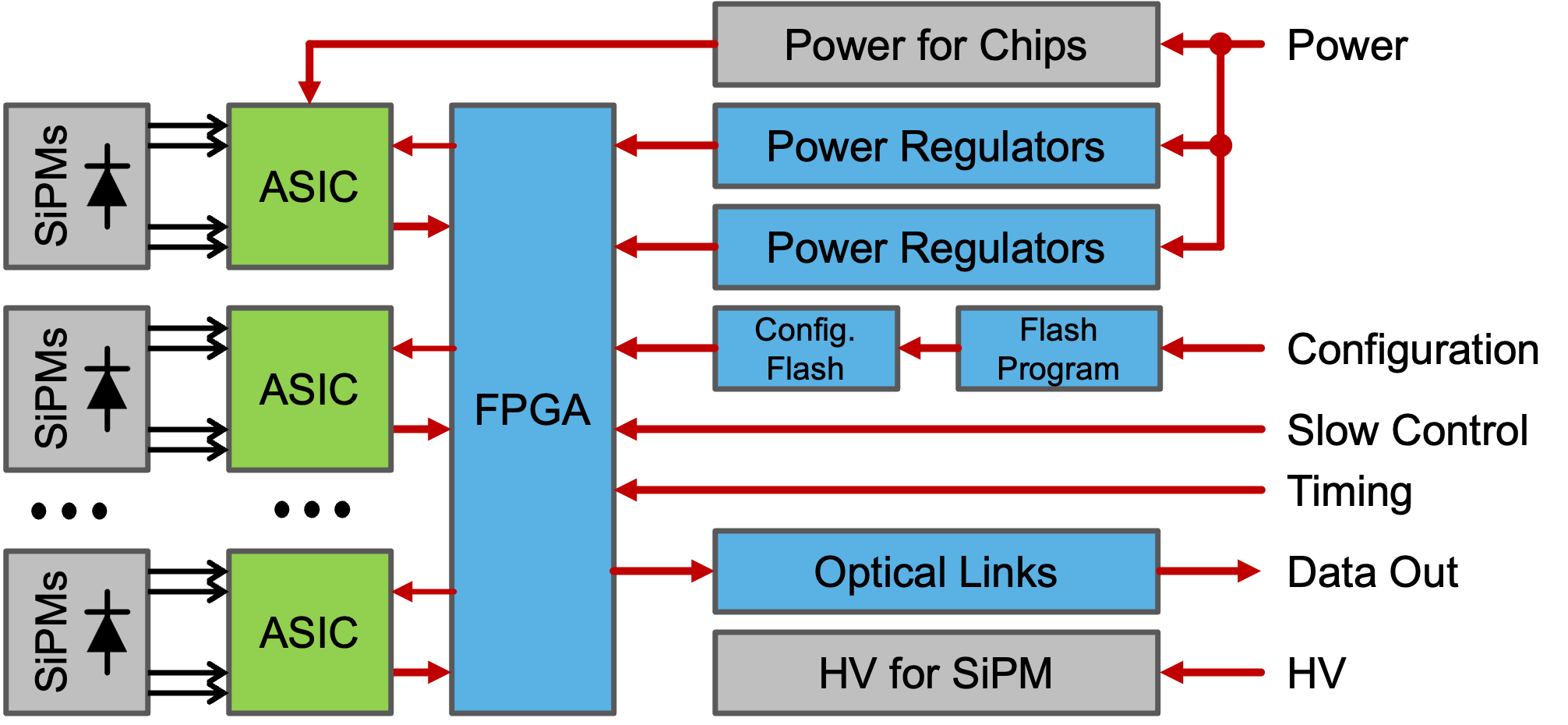}{fig:classicalsystem}{Many PET systems use readout ASICs which individually communicate with an FPGA for control and data aggregation. This requires a lot of additional auxiliary circuitry inside the PET scanner.}{0.7}

In this paper, we propose two concepts to address the above mentioned challenges. We eliminate the local \inquotes{module} FPGAs by adding more advanced readout functionality to the front-end chips, and we propose to use a serial powering scheme of the chips to reduce supply currents. Our work is based on the PETA chip family \cite{PETA} developed in our group. We denote our latest chip discussed here \inquotes{\PETAT} where \inquotes{T} stands for \inquotes{Total Body}, i.e. a scanner with many chips.

The paper is divided in two main sections, the first describing the readout concept, the second discussing the serial powering approach.

\section{Time Sorted Daisy Chain Readout}

The concept of a daisy chain readout is illustrated in \fig{fig:readoutconcept}. Each \PETAT chip processes a number of SiPM sensor signals. The hit data from these channels (times, amplitudes, channel ID, \dots) is merged and sent out on (at least) one serial {\em output} link. Each chip has in addition a number of serial {\em inputs} links which can receive and decode the output link data format. The data from the input links is merged with the data generated by the chip itself and passed \inquotes{downstream} to further chips, so that a large number of chips can be read out without any additional hardware. Depending of the number of input links per chip, different topologies are possible, from a simple linear chain of chips to fully balanced trees (see later). Obviously, the payload data rate at the end of one such readout tree depends on the PET activity and on the number of chips in the tree, which may therefore not be chosen too high. As will be explained in the next section, it is rather simple with this data aggregation approach to produce a time sorted data stream, which can simplify a lot the further hit processing.

\Figure{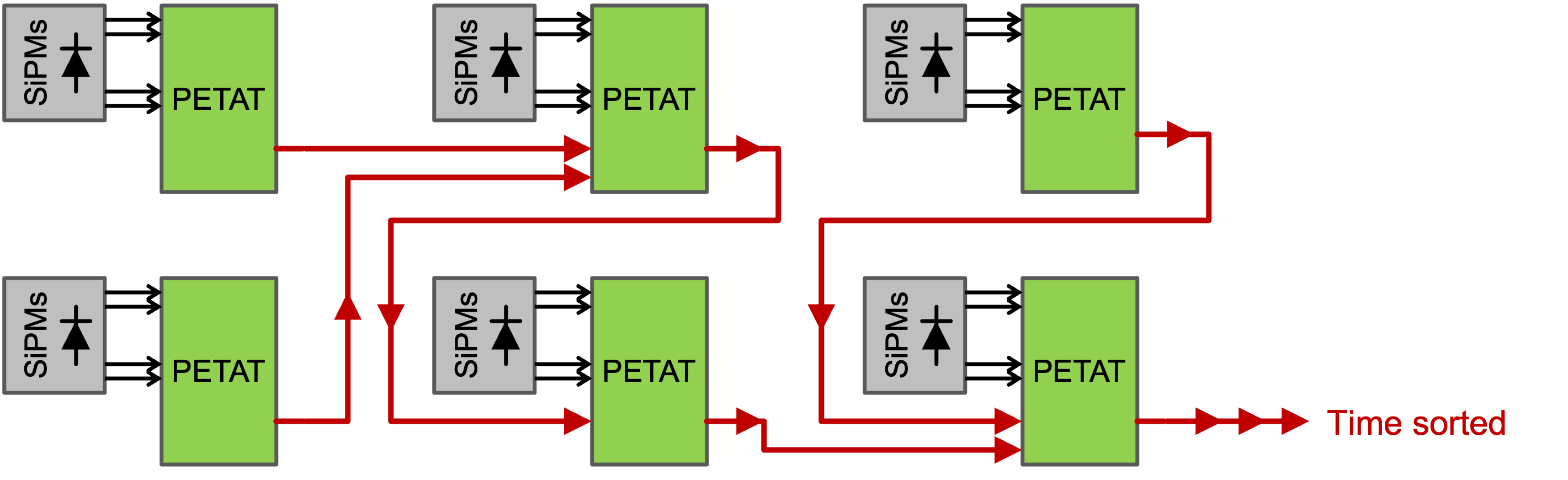}{fig:readoutconcept}{The \PETAT chips process signals from a number of SiPM sensors and send data off-chip on a serial link. Additionally, each chip can receive serial data from other chips and merge it into its output data stream. This allows aggregation of the data from several chips without additional hardware. As explained in the text, this approach makes it rather easy to obtain a time-sorted data stream.}{0.85}

\subsection{Implementation}

The most important internal components of one chip are shown in \fig{fig:blockdiagram}. A number of $N$ input links (top left) receive serial streams of hit data. The links in \PETAT use the well-known 8B10B protocol. When no data is present, \inquotes{idle} control words are received and discarded by the Ser/Par logic. A hit data packet is flagged by a different word type announcing to the Ser/Par logic that valid data will arrive. A fixed number of data words containing as payload the data of one hit is then received, parallelized and stored into a FIFO buffer. Each hit contains, among others, the ID of the chip which originally generated the hit, the SiPM channel within that chip, the hit amplitude and a time stamp indicating when the hit has occurred. The time stamp is sent with a limited number of bits, so that the count \inquotes{wraps around} after a certain time period. This must be handled properly. (In \PETATa, one time bin corresponds to $\approx50\ps$ and 20 bits are used for the time stamp, so that the wrap-around period is $\approx 52\us$.)

\Figure{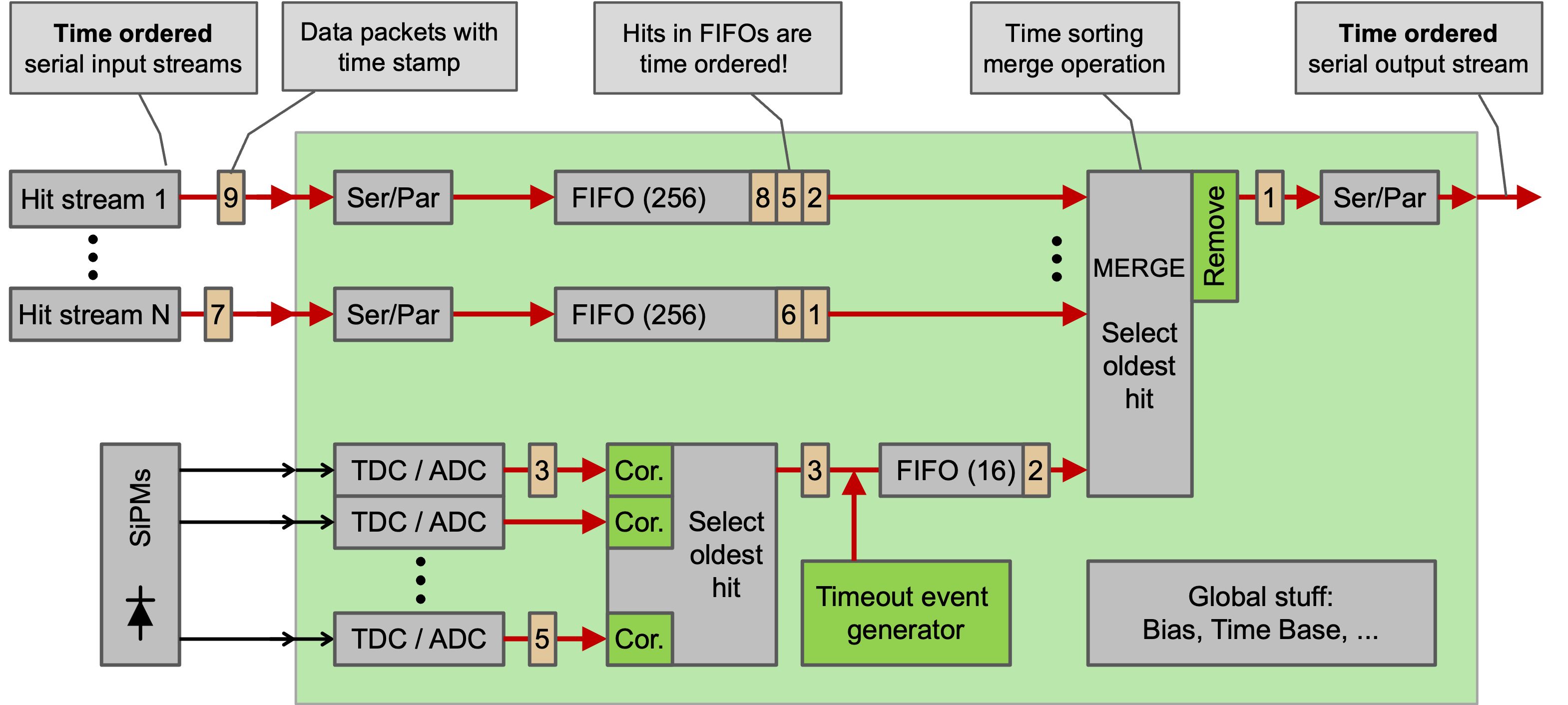}{fig:blockdiagram}{Simplified block diagram of a \PETAT chip. Serial data streams delivering time-ordered hits from other chips are parallelized and buffered in FIFOs. SiPM hits detected in the chip are digitized and time stamped (2 hits are shown), time corrected, time sorted and also buffered in a FIFO. The output data of the FIFOs is merged respecting time order, serialized and sent out. Timeout events are injected into the data stream to avoid too long waiting times in the merge step, see text.}{0.95}

We assume here that the events on each incoming link are {\em time-sorted}, i.e.\ that lower time stamps (older hits) come first. As a consequence, the hits at the {\em outputs} of the FIFOs are the oldest ones in the corresponding link. It is therefore very simple to merge the input data streams into a common, time-ordered output stream: The time stamps of all FIFO outputs are compared and the hit with the lowest value is removed from its FIFO and passed to the output link. In the (rare) case of multiple identical time stamps, an arbitration method is used. The time comparison is complicated by the wrapping-around of the time stamps. This is resolved by guaranteeing that the difference between two time stamps never exceeds half of the period, as explained below. The time-ordered output stream is the input for further chips, so that the assumption of time-ordered input data streams is self-consistent.

The above mentioned simple mechanism only works, if all FIFOs have data entries: If an empty FIFO would just be ignored, and the oldest hit of the {\em remaining} links be sent out, it could happen that an even older hit arrives late in that empty FIFO, and the output time order would be corrupted. The merger therefore has to wait until {\em all}  FIFOs have data. If the data rate in one link is particularly low (lower activity in that part of the scanner or less chips upstream), the FIFOs of the other links could fill up and eventually hits could get lost when there is no more space to write them into the FIFO\footnote{As a non negligible number of gammas is absorbed in the body and does not reach the PET ring, some readout losses smaller than this intrinsic inefficiency should not be problematic in practice.}. We solve this problem by the introduction of \inquotes{timeout events} (TE). These are generated regularly in addition to real hits with a correct time stamp (and a bit to flag them as TE) and are injected into the normal data stream. They guarantee that links are never empty and arbitration in not stalled. The TEs obviously consume link bandwidth so that their rate must be chosen as low as possible. As TEs are only needed when the rate of real events is low, there is no issue in the upstream parts of the data flow. In order to avoid an accumulation of too many TEs further downstream, they can be removed from the data flow as soon as enough real events are available. The regular occurrence of several TEs per epoch also guarantees that the wrap-around correction mentioned above works reliably. In our present chip implementation, we add sufficient TEs to ensure a data packet at least once every \textonequarter\xspace of the period. 

Besides transporting data, the chips obviously have to handle a number of SiPM sensors and thus generate hit data. Each SiPM channel contains a fast amplifier, a discriminator with programmable threshold, a time stamping circuit with $50\ps$ bin width and a 9 bit ADC. These circuit blocks are inherited from previous PETA chips. As shown in \fig{fig:readoutconcept}, a selection logic transfers the asynchronously occurring hits according to their age into a further (shorter) FIFO which is then treated in the same way as the serial input links. In our present implementation, the TEs are injected here.

The discussion so far assumes that the clock counters in all chips (delivering the time stamps) run perfectly synchronously. This is hard to achieve in a large PET scanner due to propagation delays of the time critical electrical signals like clock and reset. A correction of time stamps is therefore mandatory. An elegant approach does this immediately after time stamping, before hits are injected into the readout flow, as indicated in \fig{fig:readoutconcept}. Other sources of time shifts, like photon flight time or comparator time-walk, must be corrected elsewhere.

\begin{figure}[!t]
 \centering
 \includegraphics[width=0.48\linewidth]{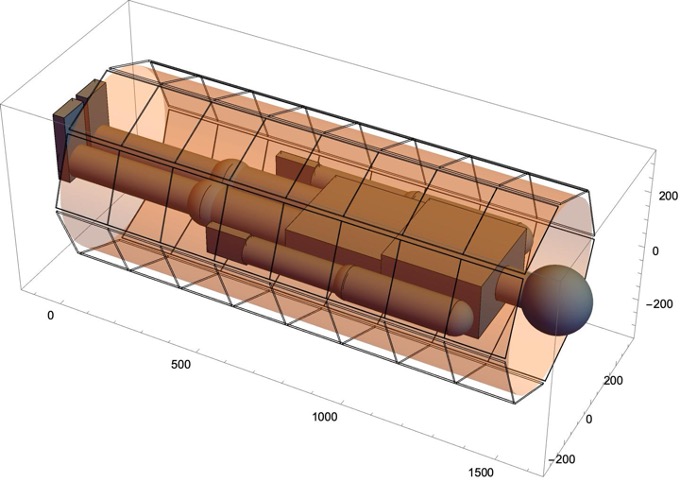}
 \includegraphics[width=0.48\linewidth]{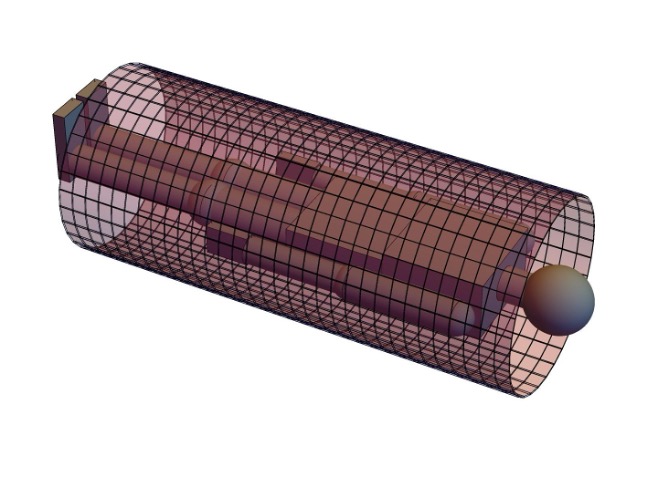}
 \vskip-2mm
 \caption{Simplified geometry used for system simulation. The cylindrical scanner of $r=30\cm$ and $l=150\cm$ is covered by a number of regions each handled by one \PETAT chip with 32 SiPM channels.} 
 \label{fig:geometry}
\end{figure}

\subsection{System Simulation}

The readout concept and chip architecture described above have been simulated for a very simplified PET scenario. A cylindrical \inquotes{total-body scanner} as shown in \fig{fig:geometry} is segmented into a larger number of regions which are each handled by one (virtual) \PETAT chip with 32 SiPM channels, respectively. We have mainly used a setup with $8\times 8$ chips, i.e.\ only 2048 SiPM channels, but the computational effort for a larger setup with $32\times 32$ chips, i.e.\ 32768 SiPMs could also be handled.

Idealized back-to-back gamma pairs are generated using simple Mathematica\textsuperscript{\textregistered} functions in a spherical central 'phantom' region of $r=15\cm$ with a fixed \inquotes{source} activity. No Compton scattering or gamma absorption is included. The gamma tracks crossing the cylinder produce hits in the area assigned to the virtual SiPMs. Gamma flight time is calculated and an ideal time stamp is assigned to the hit. No amplitude information is used. The hits generated in this way are then used to stimulate the digital simulation, which uses the {\em full}, cycle accurate HDL code describing the digital part of \PETAT. The simulation treats all 64-1024 chips in parallel, including the startup sequence (link initialization and synchronization not further described here) and chip configuration.

Several interconnection topologies (see also \fig{fig:readoutconcept}) have been investigated. They are graphically illustrated in \fig{fig:topologies}: In the linear chain (A), data is passed from one chip to the next, using only one serial input. The ideal balanced tree (B) successively merges chip pairs, using two inputs. A combination of both concepts merges chip groups in a binary tree, and connects several of such groups in a linear chain (C). Such a topology may be useful in a practical application where a number of chips is grouped on a physical module.

\Figure{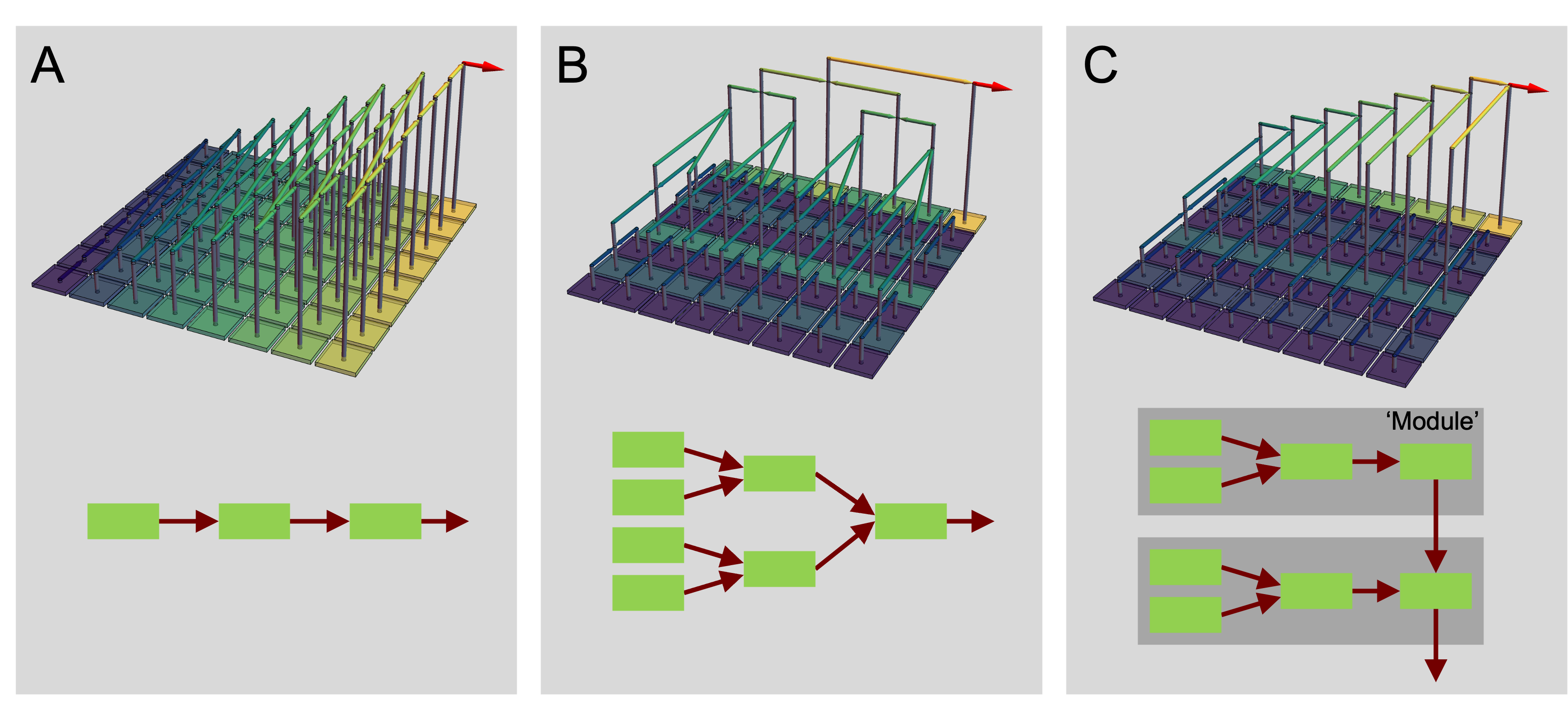}{fig:topologies}
{Different possible topologies for chip interconnection. A: linear chain, B: balanced tree, C: Groups of trees (on a \inquotes{module}) connected in a linear chain.}{0.95}

\begin{figure}[!t]
 \centering
 \includegraphics[width=0.32\linewidth]{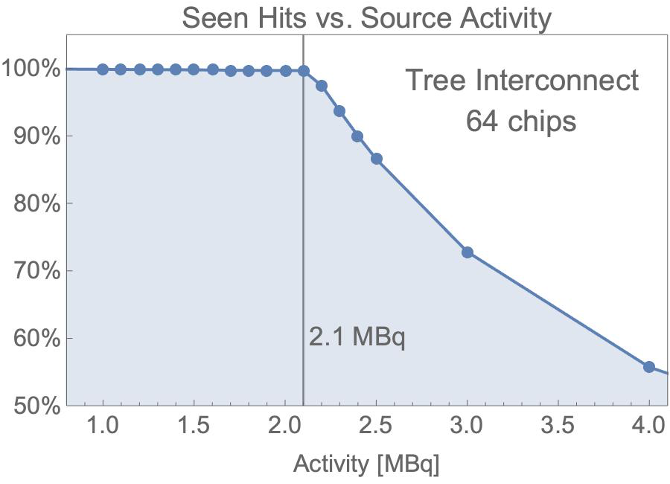}
 \includegraphics[width=0.32\linewidth]{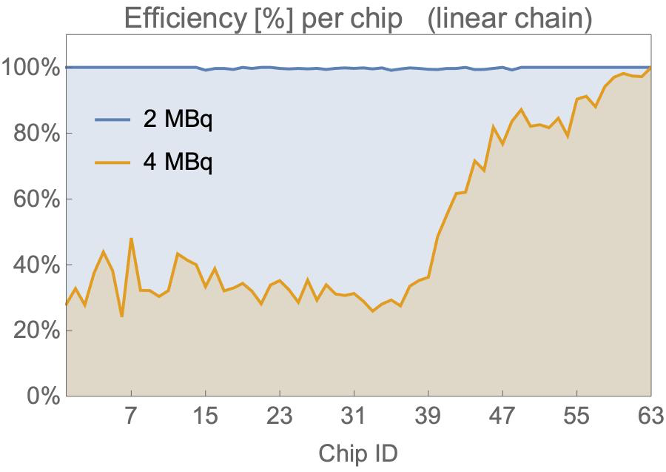}
 \includegraphics[width=0.32\linewidth]{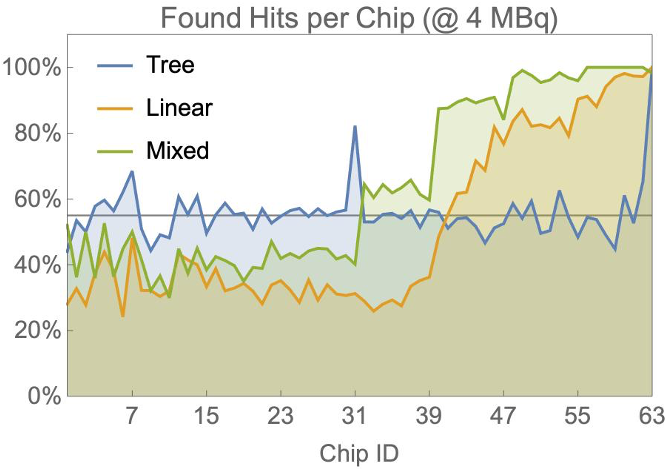}
 \vskip-2mm
 \caption{Results of cycle accurate simulation of a \inquotes{scanner} with 64 chips, see text.} 
 \label{fig:simresult}
\end{figure}

The simulation assumes a very conservative clock frequency of $312.5\MHz$ (half the frequency of the clock used for an on-chip PLL to generate time stamping bins of $\approx 50\ps$). In the present chip implementation, one hit requires 8 words of the 8B10B protocol, i.e.\ 80 bit, so that one link can transport at most $3.9\times 10^6$ hits per second. \Fig{fig:simresult} (left) shows the fraction of received hits at the end of a tree topology with 64 chips as a function of the source activity. All hits are received up to an activity of $2.1\MBq$, i.e.\ $4.2\times 10^6$ generated gammas of which some got lost at the cylinder ends. This demonstrates that the link can be efficiently filled up to its absolute bandwidth limit. \Fig{fig:simresult} (center) shows the fraction of received hits per chip for a linear chain. Nearly all hits arrive correctly for a dose of $2\MBq$, which is {\em just} below the theoretical limit. When the dose is increased above the maximal bandwidth capacity, hits (must) get lost, and the loss is unevenly distributed in the chain: Chips at the end of the chain (higher IDs) have a better chance to get their hits out, while hits from chips at the upstream part of the chain are more likely to get lost by full FIFOs. The losses (in this unrealistic overload condition) are distributed much more evenly for a tree topology, as shown in the rightmost plot. The simulated output data of the last chip has also been used to reconstruct the source volume. The emitting sphere could be reconstructed with a small degradation as expected from the finite \inquotes{crystal} size. 

\subsection{FPGA Data Processing Concept}

The hit rate per link is limited by the link clock speed (few $\GHz$ are realistic with moderate hardware effort) and the size of one data packet (addresses, time stamp, amplitude, flags) to $(10-20)\times 10^6$ hits per second. This is not sufficient for a large PET scanner so that several links must be used in parallel. Before sending their data to the acquisition system, they can be aggregated and pre-processed in {\em one} larger FPGA which can also be used to reduce the data volume by dropping useless information. \Fig{fig:fpga} shows a possible processing pipeline (which has not yet been implemented nor simulated in detail). After the serial-to-parallel conversion of each link, a last step of time sorting produces a single, time-sorted stream of unpacked hits. The time stamps can now be extended to many more bits by counting the epochs (i.e. wrap-arounds). The timeout events guarantee that no epoch can be overlooked because there are data packets from each epoch even without physical hits. The timeout events can then be dropped. In our present implementation, they also contain some monitoring information (overflow flags, chip status, temperature, $\dots$) which can be collected here. All \inquotes{singles} hits for which the time differences to the previous and to the following hit are larger than a defined coincidence time window can easily be dropped thanks to the time-ordering. Energy cuts may need clustering of hits and need more advanced processing steps, indicated in gray in \fig{fig:fpga}: After a conversion of chain/chip/channel to r/$\phi$ and an energy correction, implemented for instance by lookup tables, the total energy of a hit (if distributed over a cluster of several crystals/SiPMs) can be summed up. This is again eased by the fact that the contributing hits must have very similar time stamps and are therefore more or less grouped in the time-sorted data stream. The hit-cluster energy could be used to discard Compton scatter events, and possibly their partner hit. The data volume could further be reduced by computing the impact position of a hit cluster, for instance with a neural network approach. The remaining hits are then transferred to the data acquisition computer for further processing and reconstruction.

\Figure{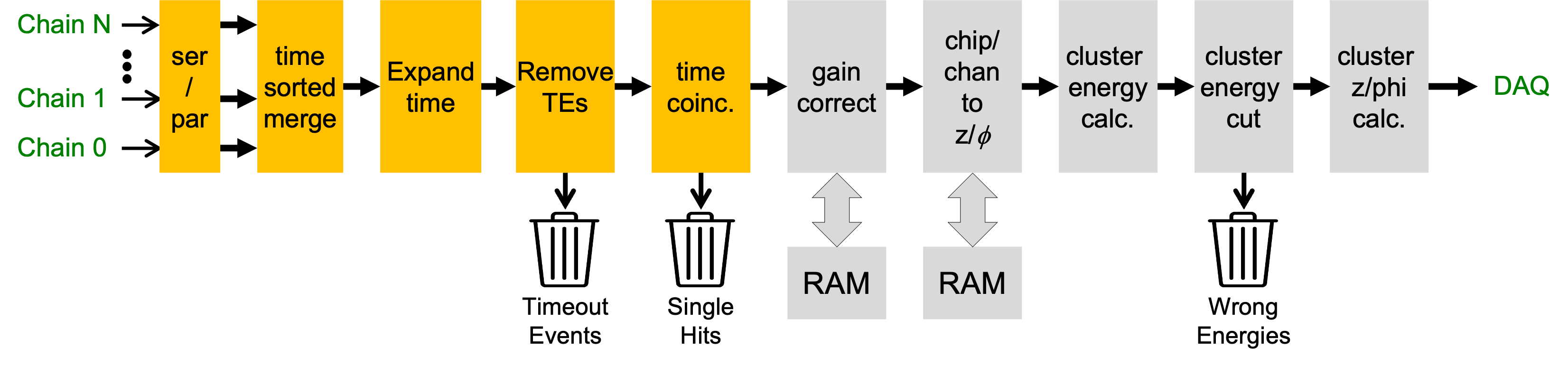}{fig:fpga}{Possible processing pipeline in a system FPGA. Advanced processing steps are shown in gray.}{0.98}

\Figure{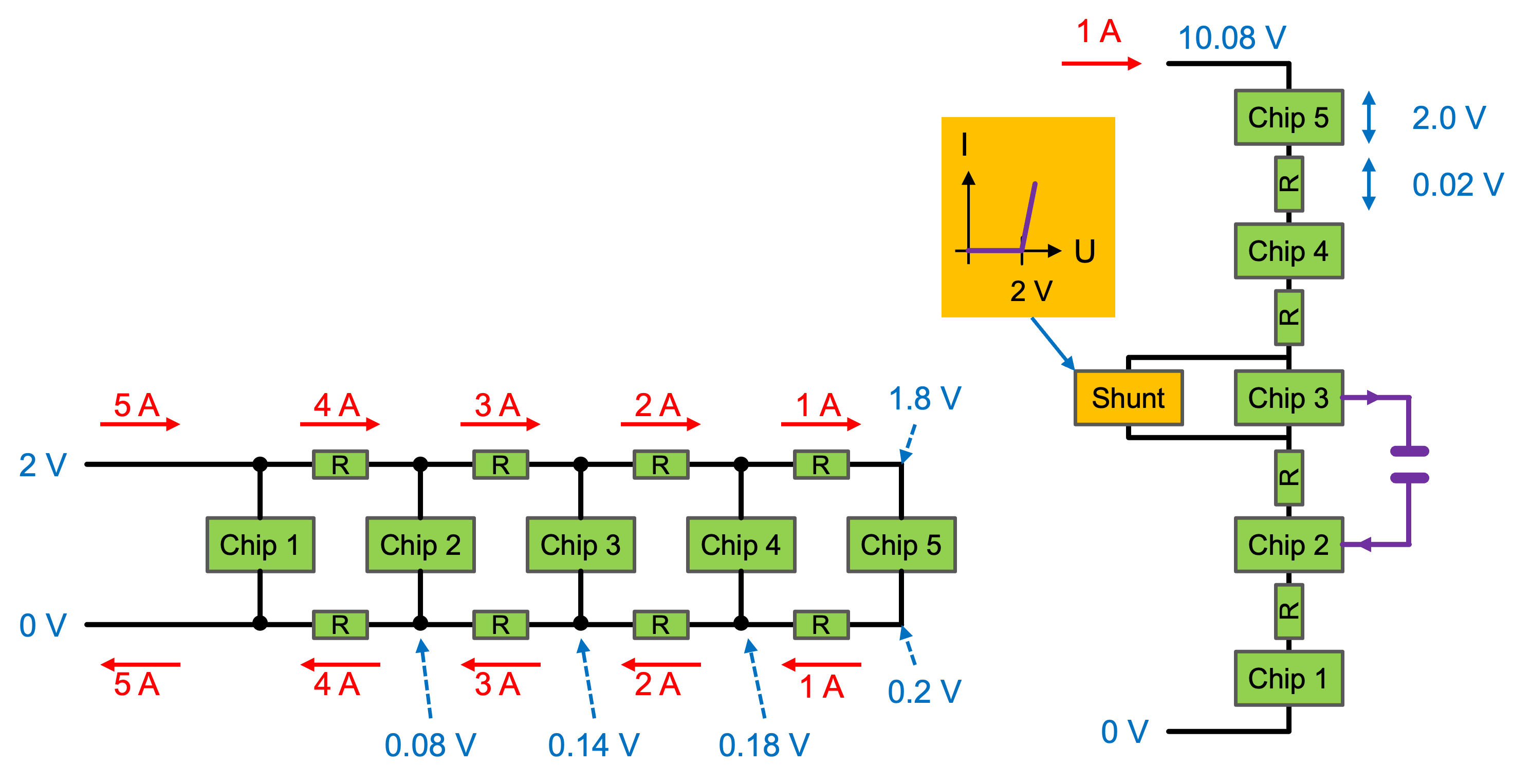}{fig:powering}{Illustration of voltage and current conditions for parallel powering (left) and serial powering (right) of chips drawing $1\Amp$ at a supply voltage of $2\V$. The resistivity of the traces/cables between chips is assumed here to be $R=0.02\Ohm$}{0.95}

\section{Serial Powering}

The high coincidence time resolution of $\approx 200\ps$ (FWHM) required in particular in ToF-PET leads to a relatively high power dissipation for each SiPM channel of $\gtrsim 5\mA$, so that the overall supply current of a larger number of PET readout chips is high. This makes the supply of stable voltages in larger systems challenging with significant losses in cables and linear voltage regulators. Similar problems in detectors for particle physics  have lead to the development of the \inquotes{serial powering} scheme, which is used now in several experiments \cite{serialpoweringATLAS,serialpoweringCMS} to reduce power losses in the modules and in the long supply cables and/or to reduce the cable cross sections. The small number of electrical connections between chips in the readout scheme proposed here makes it particularly suited for serial powering. We have therefore included circuity to evaluate this option, which we briefly describe here. 

\subsection{Powering Concept and Benefit}

The problem of resistive voltage drops when powering a larger number of chips in parallel is illustrated in \fig{fig:powering} (left) for 5 chips drawing $1\Amp$ at a nominal supply voltage of $2\V$, and assuming a chip-chip trace resistivity of $0.02\Ohm$. When powering the unit with just $2\V$, the left Chip 1 sees the correct supply voltage. The current of $4\Amp$ flowing in the power and ground traces to/from the remaining 4 chips leads to voltage drops of $2\times 0.08\V$, so that Chip 2 only sees $1.84\V$ and $0.64\W$ are dissipated in the traces. This effect repeats, to a decreasing extend, towards the other chips. The last chip in this example only sees a supply voltage of $1.6\V$. The unequal supply voltages of the chips can be unacceptable so that linear regulators with additional power dissipation may be required. The second issue is the power dissipation in the traces, which adds up to $1.2\W$, i.e.\ $12\%$ in this example. This loss grows quadratically with the number of chips! An additional source of power dissipation are the main supply cables on the left.

The situation is different in the \inquotes{serial powering} scheme where the chips are arranged in a chain, as shown in \fig{fig:powering} (right). The current flowing through the chain is $1\Amp$, ideally, so that the voltage drops in the resistors only add $0.08\V$. Assuming that each chip gets its required $2\V$, the total voltage at the chain is $10.08\V$. The power dissipated in the resistors is much lower due to the smaller current, and amounts to $0.08\W$, or $<1\%$ in this example. The dissipation in the feeding cables is reduced by a factor 5 as well. 

\Figure{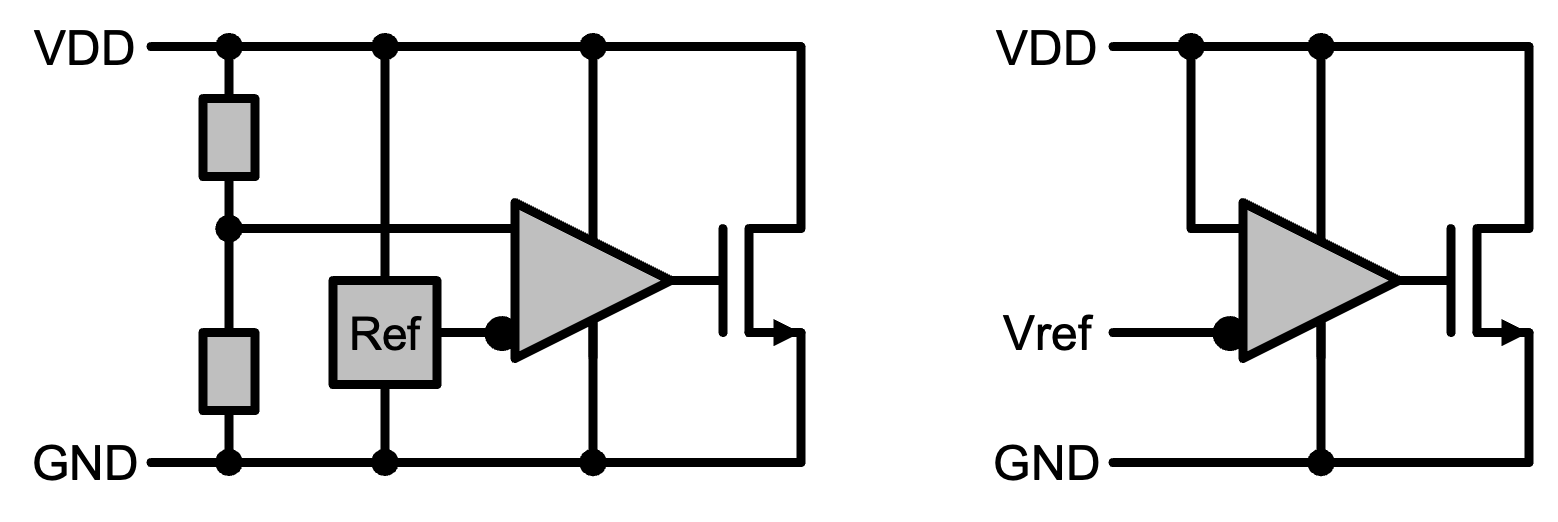}{fig:shunt}{Left: This shunt regulator compares a fraction of the chip supply voltage VDD to an internally generated reference voltage. If the supply voltage is too large, a large NMOS transistor draws current between VDD and GND. Right: A simplified version compares VDD to an externally provided reference voltage.}{0.6}

\subsection{Shunt Regulators}

The main challenge in this concept is the stabilization of the chip supply voltages to the required value. This can be accomplished by a \inquotes{shunt regulator} in parallel to each chip, which draws a very large current as soon as the chip supply voltage exceeds the nominal value. The I-U characteristic of the shunt regulator resembles that of a Zener diode and is illustrated in \fig{fig:powering}. The whole chain is then supplied with a current rather than with a voltage, which must exceed the current consumption of the worst chip in the chain. The bypass current leads to an additional power dissipation and must be minimized. \Fig{fig:shunt} shows a shunt regulator with an internal reference and the version used in 	\PETAT, which compares the positive chip supply voltage to an externally provided reference. We believe that the second version provides the voltage steps more reliably due to the absolute external reference and has better voltage stability for current transients. This is still work in progress.

\Figure{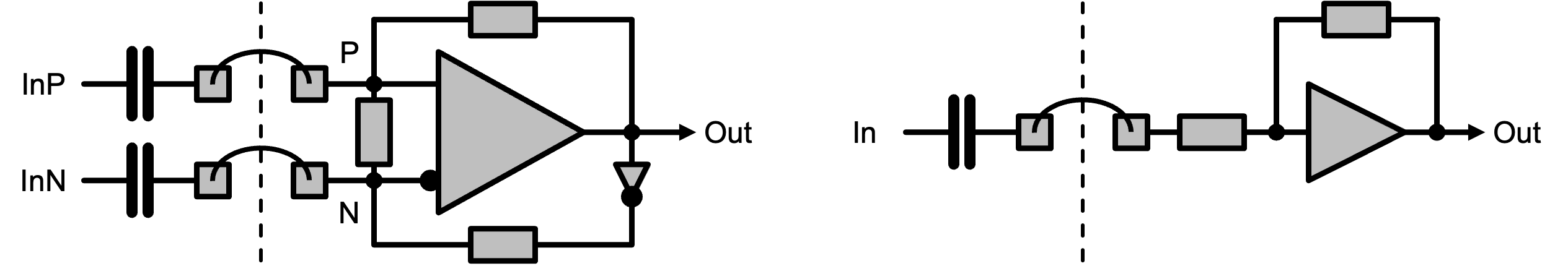}{fig:acpads}{Left/Right: AC coupled differential/single-ended input pad.}{0.95}

\subsection{Digital Links}

Due to the different ground/supply potentials of the individual chips in the serial topology, their digital output signals cannot be directly connected to other chips \inquotes{above} or \inquotes{below}. This can be solved by capacitive coupling, as indicated in \fig{fig:powering}. The circuits used in \PETAT are shown in \fig{fig:acpads}. Both use a small positive feedback to keep the input at its logical value. The input signal couples through capacitors to the storage nodes and the rising/falling edges flip the cell. As indicated, we use external capacitors for sufficient voltage robustness. In the differential cell (\fig{fig:acpads}, left), a signal $P$ more positive than $N$ leads to a logic \hex{1} at the output. This VDD potential is fed back to $P$, while a ground level is fed back to $N$ such that a permanent small positive voltage difference between $P$ and $N$ is established. A falling edge signal at $InP$ pushes $P$ low, while the simultaneous rising edge at $InN$ pushes $N$ up, so that the cell flips. Note that a \inquotes{wrong} input signal edge (at startup) would not flip the cell, the $N,P$ levels would come back to the situation described above, and \inquotes{normal} operation can begin. The single ended cell works in a similar way. Input protection diodes at the chip pads eliminate out-of-range voltage levels at startup. The digital output cells are unchanged.

The differential input cell is used for clock, the serial inputs, and a fast broadcast protocol used to reset chips and send configuration data (not further described here). The single-ended cells are uses for a standard JTAG slow control interface.

\begin{figure}[!t]
 \centering
 \includegraphics[width=0.20\linewidth]{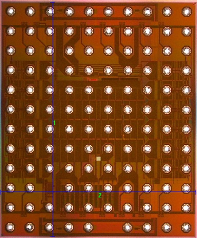}
 \includegraphics[width=0.35\linewidth]{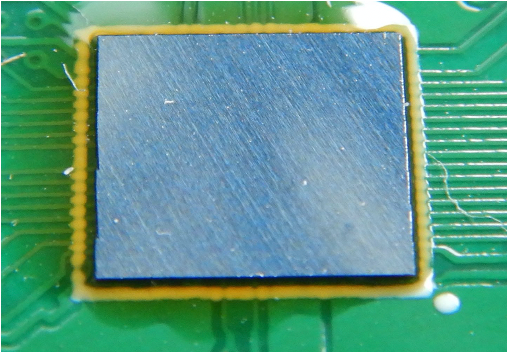}
 \includegraphics[width=0.40\linewidth]{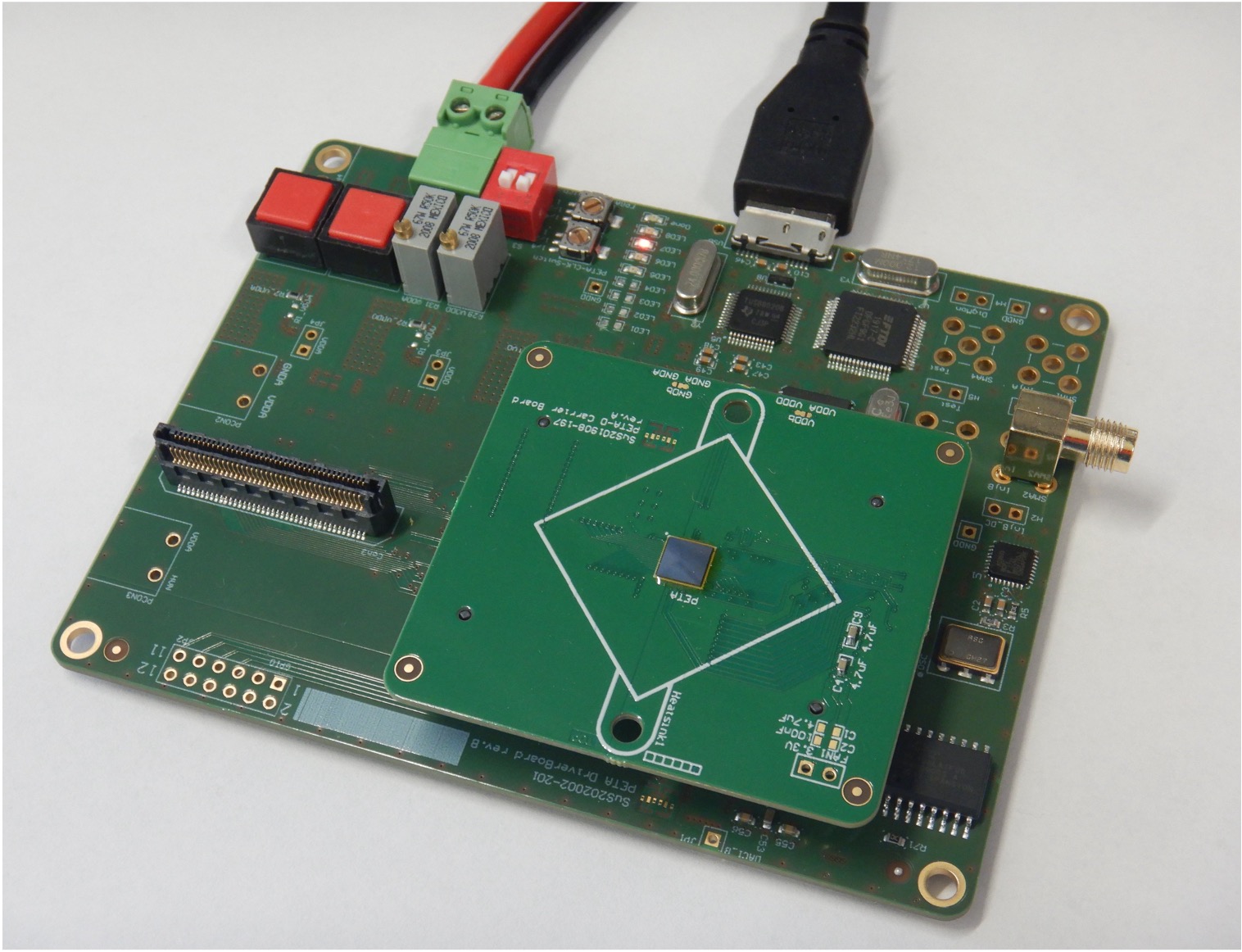}
 \vskip-2mm
 \caption{Left: Micro-photograph of the \PETATa chip with a size of $5\times 6\mm^2$. The chip uses balls in a pitch of $500\um$ for flip chip mounting. The upper and lower bump rows connect to two shunt regulators. Middle: Flipped \PETATa chip. Right: USB test setup.} 
 \label{fig:test}
\end{figure}

\section{Test Chip and Outlook}

A first chip version, \PETATa, has been designed and produced in the UMC $180\nm$ technology. It has a size of $5\times 6\mm^2$, including two shunt regulators, and uses solder balls for interconnect. \fig{fig:test} shows photographs of the chip, a flipped chip and the USB test setup developed for testing. The shunt regulators work as expected and a chain of 3 chips has been set up successfully. Chip configuration via JTAG and the fast serial transfer of data have been successfully operated across the voltage boundaries of the serially connected chips. Because the analogue SiPM section of \PETATa was configured for an experimental SiPM readout concept (block detector readout with several SiPMs being grouped in the chip to produce a common time stamp) an operation with a standard crystal array was not easily possible. 

Encouraged by our simulation results, we decided therefore to extend and improve the digital architecture of \PETATa and provide again 32 individual analogue channels, similar to the front end design that we are successfully using in the SAFIR experiment. The new chip \PETATb has been submitted early 2024 and is expected back by the end of 2024. Testing of the digital part will use the USB setup shown in \fig{fig:test} which allows to clock and synchronize the chips (with a fast broadcast signal to all chips not further described here), inject serial test data into the input links and receive and analyze the data of the output link. The SiPM readout will be tested by reading out small crystal arrays with several chips. In parallel we are working of the implementation of the FPGA data processing.



\begin{thebibliography}{00}

\bibitem{largescanner}
For instance:
S.\,Vandenberghe, P.\,Moskal and J.\,S.\,Karp:
State of the art in total body PET,
EJNMMI Physics vol. 7, Article number: 35 (2020),
DOI: 10.1186/s40658-020-00290-2

\bibitem{safirdcdc}
C.\,Ritzer et. al.:
Compact MR-compatible DC-DC converter module,
Journal of Instrumentation, Vol. 14, September 2019,
DOI 10.1088/1748-0221/14/09/P09016

\bibitem{PETA}
I.\,Sacco, P.\,Fischer and M.\,Ritzert:
PETA4: a multi-channel TDC/ADC ASIC for SiPM Readout,
Journal of Instrumentation, Vol. 8, December 2013,
DOI: 10.1088/1748-0221/8/12/c12013

\bibitem{PETSYS}
V.\,Nadig, B.\,Weissler, H.\,Radermacher, V.\,Schulz and D.\,Schug:
Investigation of the Power Consumption of the PETsys TOFPET2 ASIC,
IEEE Trans. on Rad. and Plasma Medical Sciences, Vol. 4, No. 3, May 2020,
DOI: \href{https://arxiv.org/abs/1908.05878}{1908.05878}

\bibitem{serialpoweringATLAS}
D.\,B.\,Ta et. al.:
Serial powering: Proof of principle demonstration of a scheme for the operation of a large pixel detector at the LHC,
Nucl. Instr. and Methods A, Vol. 557, Issue 2, 15 February 2006, pp. 445-459,
DOI: https://doi.org/10.1016/j.nima.2005.11.115

\bibitem{serialpoweringCMS}
R.\,Ceccarelli et. al.:
Serial powering characterisation for the CMS Inner Tracker at the High Luminosity LHC,
Journal of Instrumentation, Vol. 19, January 2024,
DOI: 10.1088/1748-0221/19/01/C01056

\bibitem{SAFIR}
J.\,Debus et. al.:
Design and first performance tests of the {SAFIR}-{II} {PET}-{MR} Scanner,
IEEE NSS, MIC and RTSD, November 2023,
DOI: 10.1109/nssmicrtsd49126.2023.10338699

\end{thebibliography}
\end{document}